\documentclass{emulateapj}

\usepackage{amsmath}
\usepackage{mathptmx}
\newcommand{\unitspace}{\ensuremath{\,}}
\newcommand{\usp}{\unitspace}
\newcommand{\numberspace}{\ensuremath{\;}}
\newcommand{\nsp}{\numberspace}
\newcommand{\unitstyle}[1]{\ensuremath{\mathrm{#1}}}
\newcommand{\power}[2]{\ensuremath{{#1}^{#2}}}

\newcommand{\centi}{\unitstyle{c}}
\newcommand{\kilo}{\unitstyle{k}}

\newcommand{\meter}{\unitstyle{m}}
\newcommand{\gram}{\unitstyle{g}}
\newcommand{\second}{\unitstyle{s}}

\newcommand{\cm}{\centi\meter}
\newcommand{\Kelvin}{\unitstyle{K}}
\newcommand{\K}{\Kelvin} 

\newcommand{\grampercc}{\gram\usp\power{\cm}{-3}} 

\newcommand{\yr}{\unitstyle{yr}}        
\newcommand{\km}{\kilo\meter}   

\newcommand{\NA}{\ensuremath{N_\mathrm{\!A}}} 

\newcommand{\ee}[1]{\ensuremath{\times 10^{#1}}}

\newcommand{\code}[1]{\textsc{#1}}

\newcommand{\nuclei}[2]{\ensuremath{\mathrm{^{#1}#2}}}

\newcommand{\neutron}{\nuclei{}{n}}
\newcommand{\nt}{\neutron}
\newcommand{\proton}{\nuclei{}{p}}
\newcommand{\pt}{\proton}

\newcommand{\helium}[1][4]{\nuclei{#1}{He}}

\newcommand{\boron}[1][11]{\nuclei{#1}{B}}
\newcommand{\carbon}[1][12]{\nuclei{#1}{C}}
\newcommand{\nitrogen}[1][14]{\nuclei{#1}{N}}
\newcommand{\oxygen}[1][16]{\nuclei{#1}{O}}
\newcommand{\fluorine}[1][19]{\nuclei{#1}{F}}
\newcommand{\neon}[1][20]{\nuclei{#1}{Ne}}
\newcommand{\sodium}[1][23]{\nuclei{#1}{Na}}
\newcommand{\magnesium}[1][24]{\nuclei{#1}{Mg}}

\newcommand{\nickel}[1][58]{\nuclei{#1}{Ni}}

\newcommand{\germanium}[1][74]{\nuclei{#1}{Ge}}

\newcommand{\SNeIa}{SNe~Ia}

\newcommand{\Slam}{\ensuremath{S_{\mathrm{lam}}}}
\newcommand{\Dlam}{\ensuremath{\delta_{\mathrm{lam}}}}
\newcommand{\vcond}{\ensuremath{\Slam}}
\newcommand{\Sdot}{\ensuremath{\varepsilon}}
\newcommand{\Atwood}{\ensuremath{\mathrm{At}}}
\newcommand{\gibson}{\ensuremath{\ell_\mathrm{G}}}

\begin{document}
\shorttitle{Laminar Flame Speedup by \protect\neon[22] Enrichment}
\shortauthors{Chamulak et al.}
\title{The Laminar Flame Speedup by Neon-22 Enrichment in White Dwarf Supernovae}
\author{David A. Chamulak, Edward F. Brown\altaffilmark{1}}
\affil{Department of Physics and Astronomy and the Joint Institute for
Nuclear Astrophysics, Michigan State University, East Lansing, MI 48824}
\and \author{Francis X. Timmes}
\affil{Thermonuclear Applications, X-2, Los Alamos National Laboratory}
\altaffiltext{1}{National Superconducting Cyclotron Laboratory, Michigan State University, East Lansing, MI 48824}
\email{chamulak@pa.msu.edu}
\email{ebrown@pa.msu.edu}
\email{timmes@lanl.gov}

\begin{abstract}
Carbon-oxygen white dwarfs contain \neon[22] formed from $\alpha$-captures onto \nitrogen\ during core He burning in the progenitor star. In a white dwarf (type Ia) supernova, the \neon[22] abundance determines, in part, the neutron-to-proton ratio and hence the abundance of radioactive \nickel[56] that powers the lightcurve.  The \neon[22] abundance also changes the burning rate and hence the laminar flame speed. We tabulate the flame speedup for different initial \carbon\ and \neon[22] abundances and for a range of densities. This increase in the laminar flame speed---about 30\% for a \neon[22] mass fraction of 6\%---affects the deflagration just after ignition near the center of the white dwarf, where the laminar speed of the flame dominates over the buoyant rise, and in regions of lower density $\sim 10^{7}\nsp\grampercc$ where a transition to distributed burning is conjectured to occur. The increase in flame speed will decrease the density of any transition to distributed burning.
\end{abstract}
\accepted{by The Astrophysical Journal Letters}
\slugcomment{to appear in the Astrophysical Journal Letters}

\keywords{nuclear reactions, nucleosynthesis, abundances --- supernovae: general --- white dwarfs --- galaxies: evolution}

\section{Introduction}\label{sec:introduction}
\label{s:introduction}

In the past decade, type Ia supernovae (hereafter, \SNeIa) have become
the premier standard candle for measuring the geometry of the universe. Although many details of the explosion are not well understood, there is a general belief that the explosion is the
thermonuclear incineration of a C/O white dwarf that has increased in
mass through accretion to just below the Chandrasekhar limit
\citep[for a review, see][]{hillebrandt.niemeyer:type}. Current observations are sampling the \SNeIa\ population out to
$z\approx 1.6$ \citep{riess.strolger.ea:type}, and future missions
will push this limit even farther to $z\lesssim 2$. The larger sample
of \SNeIa\ carries with it the prospect for discovering the
progenitors of these events and their evolution towards
ignition. Numerical models (for a sampling of recent work,
see \citealt{gamezo.khokhlov.ea:deflagrations,plewa.calder.ea:type}; \citealt{Ropke2005Type-Ia-superno})
are steadily becoming more refined and can begin to probe the
connection between the properties of the progenitor white dwarf---its
birth mass, composition, and binary companion---and the outcome of the
explosion.

The composition of the progenitor white dwarf should play a role in
setting the peak brightness and the composition of the ejecta. The C:O ratio is set by the mass of the progenitor main-seqence star, although \citet{r-opke.hillebrandt:case}
find that the C:O ratio is of secondary importance in setting the explosion energetics. After \carbon\ and \oxygen, the next most abundant nuclide is \neon[22], which is synthesized via
$\nitrogen(\alpha,\gamma)\fluorine(\beta+)\oxygen[18](\alpha,\gamma)\neon[22]$
during core He burning. The abundance of \neon[22] is therefore
proportional to the initial CNO abundance of the progenitor main
sequence star. \citet{timmes.brown.ea:variations} showed that the mass
of \nickel[56] synthesized depends \emph{linearly} on the abundance of
\neon[22] at densities where electron capture rates are much slower
than the explosion timescale, $\sim 1\nsp\second$. Simulations with
embedded tracer particles \citep{travaglio.hillebrandt.ea:nucleosynthesis,brown.calder.ea:type,Ropke2005Type-Ia-superno} have confirmed this dependance. 

These simulations studied the effect of adding \neon[22] by post-processing the $(\rho,T)$ traces, and as a result did not account for variations in either the progenitor structure or the sub-grid flame model caused by changes in the \neon[22] abundance. One-dimensional studies that did attempt to incorporate different progenitors self-consistently \citep{hoeflich.wheeler.ea:type,domnguez.h-oflich.ea:constraints} found
a much smaller dependence of the \nickel[56] yield on metallicity\footnote{It is unclear whether these studies allowed $[\mathrm{O}/{\mathrm{Fe}}]$ to vary as a function of $[\mathrm{Fe}/{\mathrm{H}}]$.}. 

The possibility that type Ia supernovae might evolve with the abundance of $\alpha$-elements in the host population, combined with questions about whether this introduces systematic variations in the Phillips relation, motivates further investigation of how the progenitor composition influences the explosion.
As a first step, we investigate in this \emph{Letter} how the abundance of \neon[22] affects
the laminar flame speed \Slam\ and width \Dlam\ of a \carbon-\oxygen-\neon[22] mixture. Our principal conclusion is that \Slam\ increases roughly \emph{linearly} with the \neon[22] mass fraction $X_{22}$. At $X_{22}=0.06$, the speedup varies, for carbon mass fraction $X_{12}=0.5$, from $\approx 30\%$ at densities $\gtrsim 5.0\ee{8}\nsp\grampercc$ to $\approx 60\%$ at lower densities.  

These calculations are relevant for two regimes: 1) the initial burn near the center of the white dwarf where the gravitational acceleration is small and the laminar flame speed dominates the evolution of a bubble of ignited material \citep[see, e.g.,][]{Zingale2006Propagation-of-}, and 2) the burning at densities $\sim 10^{7}\nsp\grampercc$ where the Gibson length scale becomes $\gibson \sim \Dlam$. The Gibson
scale $\gibson$ is defined by $v(\gibson) = \Slam$ where $v(\ell)$ is the eddy velocity for a lengthscale $\ell$ \citep[see][for a succinct review]{hillebrandt.niemeyer:type}.  The region where $\Dlam = \gibson$ is conjectured to be a possible location for a deflagration-to-detonation transition
\citep{Niemeyer1997The-Thermonucle}.  Our calculation does not apply in the flamelet regime, where the buoyancy of the hot ashes generates turbulence via the Rayleigh-Taylor instability. In this regime,  the effective front speed becomes independent of the laminar flame speed \citep{khokhlov:propagation,reinecke.hillebrandt.ea:new,Zhang2006On-the-Evolutio}, and the composition affects the front speed only through the Atwood number, $\Atwood\equiv (\rho_{\mathrm{fuel}}-\rho_{\mathrm{ash}})/(\rho_{\mathrm{fuel}}+\rho_{\mathrm{ash}})$, where $\rho_{\mathrm{fuel(ash)}}$ is the density in the unburned (burned) material.

In \S~\ref{s:numerical-results} we describe our computational method and benchmark our calculations against earlier results of \citet{timmes92}. Section \ref{s:results} presents the computed flame speeds as functions of $\rho_{\mathrm{fuel}}$, $X_{12}$, and $X_{22}$. We provide a fitted expression for \Slam\ as a function of these parameters.  We also give a physical explanation for the speedup before concluding, in \S~\ref{s:discussion}, with a discussion of how the transition to distributed burning would occur at a lower density if the \neon[22] abundance were increased.

\section{The laminar flame}\label{s:numerical-results}

To solve for the conductive flame speed \vcond, we used the assumption of isobaric conditions to cast the equation for the energy as two coupled equations for the temperature and flux \citep{timmes92,bildsten95:_propag},
\begin{eqnarray}
    \label{eq:T}
    \frac{dT}{dx} &=& - \frac{F}{K}\\
    \frac{dF}{dx} &=& \rho \Sdot - \vcond \frac{\rho C_P }{K} F.
    \label{eq:F}
\end{eqnarray}
Here $F$ is the heat flux and $C_{P}$ is the specific heat\footnote{We neglect here terms such as $\partial E/\partial X_{i}$, which account for the change in the thermal properties as the abundance of nuclide $i$ changes. These terms are much smaller than \Sdot\ for matter not in NSE.}. The heating rate \Sdot\ is given by
\begin{equation}\label{eq:Sdot}
\Sdot = \NA\sum_{i}B_{i}\frac{dY_{i}}{dt},
\end{equation}
where $B_{i}$ and $Y_{i}$ are the binding energy and abundance of
species $i$, $\NA$ is Avogadro's constant, and $d/dt = \Slam(d/dx)$.  Our reaction network incorporated 430 nuclides from n to \germanium[76] (Table~\ref{t:network}).

\begin{deluxetable}{crcrcrcr}

\tablecaption{430-Nuclide Reaction Network}
\tablehead{
\colhead{El.} & \colhead{$A$} & \colhead{El.} & \colhead{$A$} & \colhead{El.} & \colhead{$A$} & \colhead{El.} & \colhead{$A$} }
\startdata
  n & & & & & & & \\
  H & 1--3      &   F &  15--24 & Cl &  31--44 & Mn &  46--63 \\
 He &   3--4    & Ne &  17--28 & Ar &  31--47 & Fe &  46--66 \\
 Li &   6--8    & Na &  20--31 &  K &  35--46 & Co &  50--67 \\
 Be & 7, 9--11  & Mg &  20--33 & Ca &  35--53 & Ni &  50--73 \\
  B &  8, 10-14 & Al &  22--35 & Sc &  40--53 & Cu &  56--72 \\
  C &  9--16    & Si &  22--38 & Ti &  39--55 & Zn &  55--72 \\
  N &  12--20   & P &  26--40 &  V &  43--57 & Ga &  60--75 \\
  O &  13--20   & S &  27--42 & Cr &  43--60 & Ge &  59--76 \\
\enddata
\label{t:network}
\end{deluxetable}

We use the reaction rates from the library \code{reaclib}
\citep[][and references therein]{rauscher00,Sakharuk2006An-Updated-Libr}. On the timescale of the flame
passage, electron captures are unimportant, and $Y_{e}$ is essentially fixed; we found that \Slam\ was unchanged when weak reactions were removed from the network, so we used only strong rates for computational efficiency.
We incorporated screening using the formalism of \citet{graboske.dewitt.ea:screening}.  Across the flame front, the matter does not reach nuclear statistical equilibrium or quasi-nuclear statistical equilibrium until most of the \carbon\ is depleted. 
Thus, although our treatment of screening does not preserve detailed balance \citep{Calder2006Capturing-the-f} this does not affect our calculation of the flame speed. 
Our equation of state has contributions from electrons, radiation, and strongly coupled ions. We include thermal transport by both degenerate electrons and photons (for a complete description of our thermal routines, see \citealt{brown.bildsten.ea:variability}, and references therein).

Equations~(\ref{eq:T})--(\ref{eq:Sdot}), when combined with appropriate
boundary conditions, have \vcond\ as an eigenvalue.  Ahead of the flame the
material is at an arbitrary cold temperature $T_{\mathrm{fuel}} = 10^8 K$; in this region we set
$dT/dx$ to a small positive value and integrate equations (\ref{eq:T})--(\ref{eq:Sdot}). For simplicity, we split the solution of the thermal and network equations; that is, for each step $dx$ we solve the thermal equations,  to obtain $T$ and $\rho$, integrate the reaction network at that $T$ and $\rho$ to compute $Y_{i}$ and \Sdot, and use that to advance the solution of the thermal equations. For our choice of \vcond, the second boundary condition is that $F\to 0$ asymptotically behind the front and that $F$ is peaked where $\Sdot $ is maximum.  We iterated until \vcond\ had converged to within 0.01\%.  We find that \Slam\ is insensitive to $T_{\mathrm{fuel}}$ for $\rho_{\mathrm{fuel}}\gtrsim 5\ee{8}\nsp\grampercc$. At lower densities this is no longer true, but the relative increase in \Slam\ with $X_{22}$ remains robust.

\begin{figure}[htbp]
\centering{\includegraphics[width=3.4in]{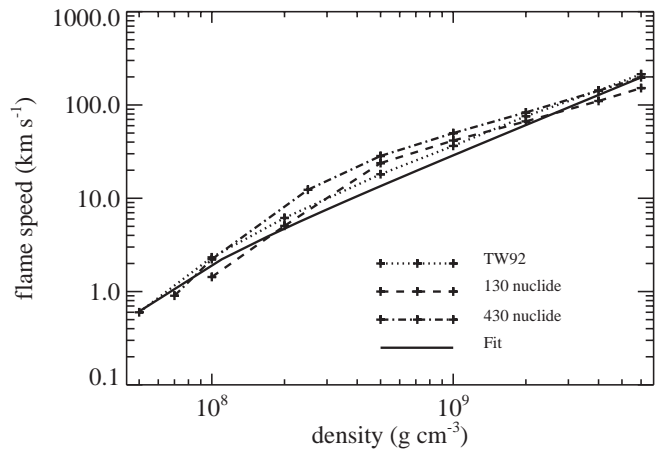}}
\caption{Flame speeds computed with an 130-nuclide network (\emph{dashed line}) and a 430-nuclide network (\emph{dash-dotted line}). We compare these with the results of \citet[\emph{dotted line}]{timmes92}, and our fit (eq.~[\ref{e:speed_fit}]; \emph{solid line}). Our 130-nuclide network uses the same nuclides as \protect\citet{timmes92}.}
\label{f:VvsD}
\end{figure}

Figure~\ref{f:VvsD} shows a comparison between our flame speeds for a 1:1 C:O mixture and those of \citet{timmes92}.  Here we adopt the same 130-nuclides as used in that paper.
Although in this case we are using the same nuclides and starting composition, the rates,
equation of state, and thermal conductivities are not identical. Overall, our flame speeds differ by no more than 25\%; the largest discrepancy is at $\rho_{\mathrm{fuel}} = 10^8\nsp\grampercc$. Most of this discrepancy is due to different opacities used by the two codes. For $\rho_{\mathrm{fuel}}\lesssim 7\ee{8}\nsp\grampercc$, photons become more efficient  than electrons at transporting heat within the flame front, with the dominant opacity being free-free. \citet{timmes92} included a fit to electron-ion scattering in the semi-degenerate regime  \citep{Iben1975Thermal-pulses;}. At these densities where $T> 2\ee{9}\nsp\K$ and the free-free opacity dominates, the contribution from electron-ion scattering decreases the total opacity. We compared our opacities along a $(\rho,T)$ trace generated for a run at $\rho_{\mathrm{fuel}} = 10^8\nsp\grampercc$.  We found that non-degenerate electron-ion scattering can lower the opacity by $\approx 24\%$, depending on how the opacity is interpolated between the two limiting fits. In addition, our free-free opacities differ by 30\% at the location along the $(\rho,T)$ trace where $|F|$ is maximum.

Finally, we also investigated the effect of reaction network size:  increasing the network from 130 to 430 nuclides resulted in a 25\% increase in \Slam\ at $\rho_{\mathrm{fuel}} = 2.0\ee{9}\nsp\grampercc$; further increases in the size of the reaction network did not yield any appreciable increases in \Slam. 

\section{Results}
\label{s:results}

We now present the results of our flame calculations for different initial mixtures of \carbon,
\oxygen, and \neon[22] and different ambient densities.
Table~\ref{t:calculations} lists \Slam\ and the flame width defined by $\Dlam = (T_{\mathrm{ash}}-T_{\mathrm{fuel}})/\max |dT/dx|$, with $T_{\mathrm{fuel(ash)}}$ being the temperature in the unburned (burned) matter.  We tabulate these quantities for $\rho_{9} \equiv \rho_{\mathrm{fuel}}/10^{9}\nsp\grampercc$ ranging from 0.05 to 6.0, and $X_{12} = 0.3\textrm{--}0.7$, with the remaining composition being \oxygen\ and \neon[22].  For each choice of $\rho_{9}$ and $X_{12}$, we use 3 different \neon[22] abundances, $X_{22} = 0.0$, 0.02, and 0.06. Over most of the range in $\rho_{9}$, $X_{12}$, and $X_{22}$ in Table~\ref{t:calculations}, we find that an increase in $X_{22}$ from 0 to 0.06 causes \vcond\ to increase by approximately 30\%. 
We confirmed several of the table entries using an independent diffusion equation solver that uses
adaptive grids \citep{timmes92} and a different reaction network and opacity routine. From this, we estimate that the uncertainty in the flame speeds listed in Table 2 are $\approx  30\%$, with about 10\% coming from numerical uncertainty and about 20\% from different physics treatments as described above. We emphasize, however, that both codes find the same trends, e.g., an increase in \Slam\ with $X_{22}$.

We fit the tabulated \Slam\ with the approximate expression
\begin{eqnarray}\label{e:speed_fit}
\Slam &=& \left[ 23.26 \rho_{9} + 37.34 \rho_{9}^{1.1} - 1.288\right] \times \left[1 + 0.3\left(\frac{X_{22}}{0.06}\right) \right]\nonumber\\
&&{}\times\left[ 0.3883 \left(\frac{X_{12}}{0.5}\right) + 0.09773\left(\frac{X_{12}}{0.5}\right)^{3}\right]\nsp\km\usp\second^{-1}
\end{eqnarray}
which has fit errors, as compared against speeds calculated using the 430-nuclide network with $X_{22}=0$, that average 33\%, with a maximum of 70\%, for $X_{12}=0.5$ and $\rho_{9}=0.07$, and with a minimum of 0.1\%, for $X_{12} = 0.5$ and $\rho_{9} = 6$. For accurate work, interpolation from Table~\ref{t:calculations} is preferred.  At $\rho \lesssim 10^{8}\nsp\grampercc$, the speedup is negligible for $X_{12} = 0.3$ but increases to $\approx 50\%$ for $X_{12} = 0.5$.
 
\begin{deluxetable*}{rrrcccrrrccc}
\tablecaption{Laminar flame speed and width}
\tablecolumns{12}
\tablehead{
\colhead{$X_{12}$} & \colhead{$X_{16}$} & \colhead{$X_{22}$} & \colhead{$\rho$} & \colhead{$\vcond$} & \colhead{$\Dlam$} &
\colhead{$X_{12}$} & \colhead{$X_{16}$} & \colhead{$X_{22}$} & \colhead{$\rho$} & \colhead{$\vcond$} &  \colhead{$\Dlam$} \\
& & & \colhead{$(10^{9}\nsp\grampercc)$} & \colhead{$(\km\usp\second^{-1})$} & \colhead{$(\cm)$} &
& & & \colhead{$(10^{9}\nsp\grampercc)$} & \colhead{$(\km\usp\second^{-1})$} &  \colhead{$(\cm)$} }
\startdata
1.00 & 0.00 & 0.00 & 4.00 & 380.200 & $7.7310\ee{-6}$ & 0.50 & 0.50 & 0.00 & 4.00 & 142.200 & $1.9696\ee{-5}$ \\
0.94 & 0.00 & 0.06 & 4.00 & 416.300 & $6.8872\ee{-6}$ & 0.50 & 0.48 & 0.02 & 4.00 & 157.400 & $1.7909\ee{-5}$ \\
1.00 & 0.00 & 0.00 & 2.00 & 230.000 & $1.8347\ee{-5}$ & 0.50 & 0.44 & 0.06 & 4.00 & 187.500 & $1.4526\ee{-5}$ \\
0.98 & 0.00 & 0.02 & 2.00 & 239.900 & $1.7351\ee{-5}$ & 0.50 & 0.50 & 0.00 & 2.00 & 83.320 & $4.6371\ee{-5}$ \\
0.94 & 0.00 & 0.06 & 2.00 & 257.700 & $1.5973\ee{-5}$ & 0.50 & 0.48 & 0.02 & 2.00 & 93.230 & $4.1110\ee{-5}$ \\
1.00 & 0.00 & 0.00 & 1.00 & 143.100 & $4.4374\ee{-5}$ & 0.50 & 0.44 & 0.06 & 2.00 & 113.000 & $3.3564\ee{-5}$ \\
0.98 & 0.00 & 0.02 & 1.00 & 150.500 & $4.2189\ee{-5}$ & 0.50 & 0.50 & 0.00 & 1.00 & 49.930 & $1.1517\ee{-4}$ \\
0.94 & 0.00 & 0.06 & 1.00 & 164.000 & $3.8244\ee{-5}$ & 0.50 & 0.48 & 0.02 & 1.00 & 56.510 & $1.0097\ee{-4}$ \\
1.00 & 0.00 & 0.00 & 0.50 & 88.020 & $1.0884\ee{-4}$ & 0.50 & 0.44 & 0.06 & 1.00 & 69.710 & $8.1250\ee{-5}$ \\
0.98 & 0.00 & 0.02 & 0.50 & 93.160 & $1.0244\ee{-4}$ & 0.50 & 0.50 & 0.00 & 0.50 & 28.400 & $3.0685\ee{-4}$ \\
0.94 & 0.00 & 0.06 & 0.50 & 102.700 & $9.2221\ee{-5}$ & 0.50 & 0.48 & 0.02 & 0.50 & 32.620 & $2.6820\ee{-4}$ \\
0.70 & 0.30 & 0.00 & 6.00 & 304.100 & $7.9262\ee{-6}$ & 0.50 & 0.44 & 0.06 & 0.50 & 41.090 & $2.1150\ee{-4}$ \\
0.70 & 0.28 & 0.02 & 6.00 & 329.700 & $7.1878\ee{-6}$ & 0.50 & 0.50 & 0.00 & 0.25 & 12.330 & $1.0643\ee{-3}$ \\
0.70 & 0.24 & 0.06 & 6.00 & 379.100 & $6.1304\ee{-6}$ & 0.50 & 0.48 & 0.02 & 0.25 & 14.620 & $1.0232\ee{-3}$ \\
0.70 & 0.30 & 0.00 & 4.00 & 220.900 & $1.2808\ee{-5}$ & 0.50 & 0.44 & 0.06 & 0.25  & 19.840 & $7.3774\ee{-4}$ \\
0.70 & 0.28 & 0.02 & 4.00 & 240.900 & $1.1637\ee{-5}$ & 0.50 & 0.50 & 0.00 & 0.10 & 2.219 & $9.1838\ee{-3}$ \\
0.70 & 0.24 & 0.06 & 4.00 & 279.700 & $9.8633\ee{-6}$ & 0.50 & 0.48 & 0.02 & 0.10 & 2.340 & $8.7795\ee{-3}$ \\
0.70 & 0.30 & 0.00 & 2.00 & 131.800 & $2.9978\ee{-5}$ & 0.50 & 0.44 & 0.06 & 0.10 & 3.500 & $4.8597\ee{-3}$ \\
0.70 & 0.28 & 0.02 & 2.00 & 145.200 & $2.7078\ee{-5}$ & 0.50 & 0.50 & 0.00 & 0.07 & 0.902 & $1.7421\ee{-2}$ \\
0.70 & 0.24 & 0.06 & 2.00 & 171.500 & $2.2688\ee{-5}$ & 0.50 & 0.48 & 0.02 & 0.07 & 1.079 & $1.2229\ee{-2}$ \\
0.70 & 0.30 & 0.00 & 1.00 & 81.420 & $7.2334\ee{-5}$ & 0.50 & 0.44 & 0.06 & 0.07 & 1.535 & $1.1516\ee{-2}$ \\
0.70 & 0.28 & 0.02 & 1.00 & 90.550 & $6.5244\ee{-5}$ & 0.30 & 0.70 & 0.00 & 6.00 & 124.400 & $1.8832\ee{-5}$ \\
0.70 & 0.24 & 0.06 & 1.00 & 108.600 & $5.4095\ee{-5}$ & 0.30 & 0.68 & 0.02 & 6.00 & 137.000 & $1.7063\ee{-5}$ \\
0.70 & 0.30 & 0.00 & 0.50 & 49.430 & $1.8119\ee{-4}$ & 0.30 & 0.64 & 0.06 & 6.00 & 163.000 & $1.3936\ee{-5}$ \\
0.70 & 0.28 & 0.02 & 0.50 & 55.460 & $1.6002\ee{-4}$ & 0.30 & 0.70 & 0.00 & 4.00 & 87.760 & $3.1546\ee{-5}$ \\
0.70 & 0.24 & 0.06 & 0.50 & 67.430 & $1.3231\ee{-4}$ & 0.30 & 0.68 & 0.02 & 4.00 & 97.180 & $2.8335\ee{-5}$ \\
0.70 & 0.30 & 0.00 & 0.25 & 26.850 & $4.8933\ee{-4}$ & 0.30 & 0.64 & 0.06 & 4.00 & 116.800 & $2.2986\ee{-5}$ \\
0.70 & 0.28 & 0.02 & 0.25 & 30.590 & $4.2824\ee{-4}$ & 0.30 & 0.70 & 0.00 & 2.00 & 49.440 & $7.6417\ee{-5}$ \\
0.70 & 0.24 & 0.06 & 0.25 & 34.370 & $3.5526\ee{-4}$ & 0.30 & 0.68 & 0.02 & 2.00 & 55.190 & $6.8173\ee{-5}$ \\
0.70 & 0.30 & 0.00 & 0.10 & 6.222 & $3.0697\ee{-3}$ & 0.30 & 0.64 & 0.06 & 2.00 & 66.910 & $5.6030\ee{-5}$ \\
0.70 & 0.28 & 0.02 & 0.10 & 7.988 & $2.4567\ee{-3}$ & 0.30 & 0.70 & 0.00 & 1.00 & 28.210 & $1.9627\ee{-4}$ \\
0.70 & 0.24 & 0.06 & 0.10 & 9.591 & $1.8043\ee{-3}$ & 0.30 & 0.68 & 0.02 & 1.00 & 31.360 & $1.7728\ee{-4}$ \\
0.70 & 0.30 & 0.00 & 0.07 & 3.042 & $6.9636\ee{-3}$ & 0.30 & 0.64 & 0.06 & 1.00 & 37.780 & $1.4747\ee{-4}$ \\
0.70 & 0.28 & 0.02 & 0.07 & 3.944 & $5.5058\ee{-3}$ & 0.30 & 0.70 & 0.00 & 0.50 & 14.930 & $5.6327\ee{-4}$ \\
0.70 & 0.24 & 0.06 & 0.07 & 5.389 & $4.1641\ee{-3}$ & 0.30 & 0.68 & 0.02 & 0.50 & 16.250 & $5.1951\ee{-4}$ \\
0.70 & 0.30 & 0.00 & 0.05 & 1.499 & $1.1291\ee{-2}$ & 0.30 & 0.64 & 0.06 & 0.50 & 18.380 & $4.4794\ee{-4}$ \\
0.70 & 0.28 & 0.02 & 0.05 & 1.992 & $1.1196\ee{-2}$ & 0.30 & 0.70 & 0.00 & 0.25 & 2.274 & $4.4388\ee{-3}$ \\
0.70 & 0.24 & 0.06 & 0.05 & 2.204 & $9.8265\ee{-3}$ & 0.30 & 0.68 & 0.02 & 0.25 & 3.404 & $3.1490\ee{-3}$ \\
0.50 & 0.50 & 0.00 & 6.00 & 197.800 & $1.2094\ee{-5}$ & 0.30 & 0.64 & 0.06 & 0.25 & 3.420 & $2.9321\ee{-3}$ \\
0.50 & 0.48 & 0.02 & 6.00 & 217.000 & $1.0968\ee{-5}$ & 0.30 & 0.70 & 0.00 & 0.10 & 1.202 & $1.5404\ee{-2}$ \\
0.50 & 0.44 & 0.06 & 6.00 & 255.800 & $9.0067\ee{-6}$ & 0.30 & 0.68 & 0.02 & 0.10 & 1.147 & $1.6142\ee{-2}$ \\
\enddata
\label{t:calculations}
\end{deluxetable*}%

\begin{figure}[htp]
\centering{\includegraphics[width=3.1in]{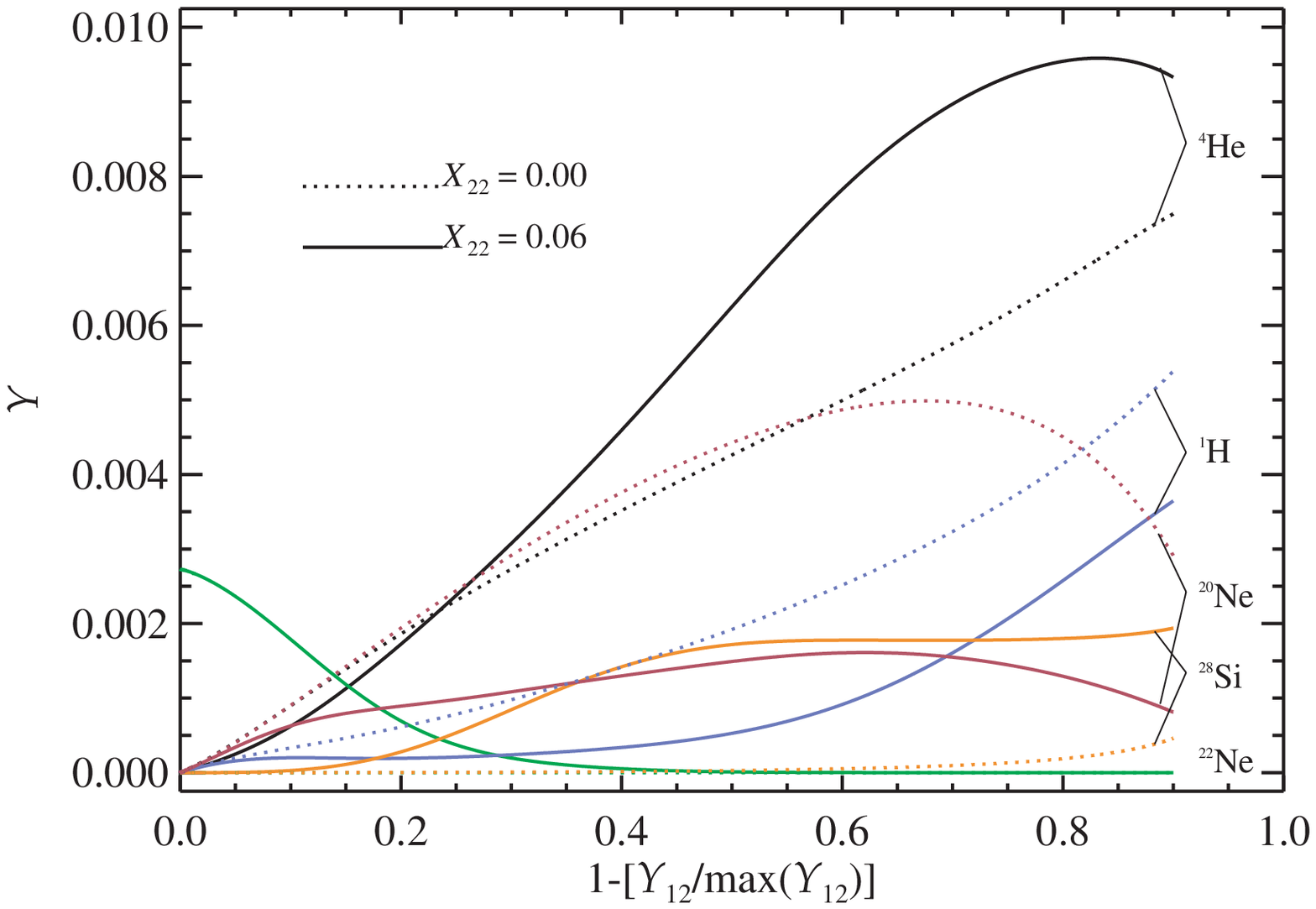}}
\caption{Abundances of selected nuclides during a burn at $\rho=2.0\ee{9}\nsp\grampercc$ and with an initial \carbon\ mass fraction of 0.3.  We show runs with an initial \neon[22] abundance of 0.06 (\emph{solid lines}) and 0.0 (\emph{dashed lines}).}
\label{f:sample-abun}
\end{figure}

To understand how the addition of \neon[22] increases \Slam, we plot in Fig.~\ref{f:sample-abun} some selected abundances $Y = X/A$ for a flame with an initial $X_{12} = 0.3$ and $\rho_{\mathrm{fuel}} = 2.0\ee{9}\nsp\grampercc$. We use the fraction of \carbon\ consumed, $1-Y_{12}/\max(Y_{12})$, as our coordinate and plot the region where this value is monotonic with distance. In a C/O deflagration, the flame speed and width are set by the initial burning of \carbon.  The buildup of Si-group nuclides and then establishment of nuclear statistical equilibrium occur on longer timescales, so that the peak of the heat flux is reached as \carbon\ is depleted via the reactions $\carbon(\carbon,\alpha)\neon$ and $\carbon(\carbon,\pt)\sodium(\pt,\alpha)\neon$.  For the case with $X_{22} = 0.06$ (\emph{solid lines}) one sees that \neon[22] is depleted before the \carbon\ is even half-consumed. 
The \neon[22] lifetime becomes less than the \carbon\ lifetime once the $\alpha$ abundance is $Y_{4} \gtrsim 10^{-4}$. At temperatures in the flame front the uncertainty in the $\neon[22](\alpha,\nt)\magnesium[25]$ rate is estimated to be about 10\% \citep[see][and references therein]{Karakas:2006sh}. This is unlike the case in AGB stars, for which the uncertainty at $T < 3\ee{8}\nsp\K$ is approximately a factor of 10.
Note that significant burning does not occur until the \carbon\ lifetime becomes of order the time for the flame front to move one flame width. This requires temperatures in excess of $2\ee{9}\nsp\K$ for the densities of interest, and so \neon[22] is preferentially destroyed in a flame via $(\alpha,\nt)$ rather than by \pt-capture \citep[cf.][]{Podsiadlowski2006Cosmological-Im}.

The neutrons made available from the destruction of \neon[22] capture
preferentially onto \neon[20] formed during \carbon\ burning. At
$\rho_{\mathrm{fuel}}\gtrsim 5\ee{8}\nsp\grampercc$, successive
$(\nt,\alpha)$ reactions build up \oxygen[17] and \carbon[14], the
latter of which then undergoes $\carbon[14](\pt,\nt)\nitrogen(\nt,\alpha)\boron[11](\pt,2\alpha)\helium$. At densities of $\rho_{\mathrm{fuel}}\lesssim 5\ee{8}\nsp\grampercc$ and carbon abundances $X_{12}=0.5$, the flow $\neon[20](\nt,\gamma)\neon[21](\nt,\alpha)\oxygen[18](\pt,\alpha)\nitrogen[15](\pt,\alpha)\carbon[12]$ also contributes. 
The net effect  of having \neon[22] in the fuel mixture is that during \carbon\ burning, the abundance of protons is depressed and the abundance of \helium\ elevated, as illustrated in Fig.~\ref{f:sample-abun}.
The fact that these flows require two neutron captures onto the products of $\carbon+\carbon$ suggests that the increase in \Sdot\ should scale roughly as $X_{22}^{2}$. Since $\Slam \propto \Sdot^{1/2}$, this implies that the increase in flame speed will be linear in $X_{22}$, which agrees with the numerical solution of equations~(\ref{eq:T})--(\ref{eq:Sdot}).  Because these flows are initiated by $\nt$-capture onto \neon, we tested our sensitivity to the reaction rate by recomputing the case $X_{12}=0.5$ and $\rho_{\mathrm{fuel}}=7.0\ee{7}\nsp\grampercc$.  A decrease in the $\neon+\nt$ rate by a factor of 10 produced a decrease in the flame speedup, from 70\% to 20\%. At higher densities there was no difference in the speedup. The only case for which there was no increase in \Slam\ was for $\rho_{\mathrm{fuel}}\lesssim 10^{8}\nsp\grampercc$ and $X_{12}=0.3$ (see Table~\ref{t:calculations}).  For this case, the slower consumption of \neon[22] relative to \carbon\ and $(\nt,\gamma)$ captures on \magnesium[25] suppress the generation of $\alpha$-particles  early in the burn.

\section{Discussion}
\label{s:discussion}

We have computed the laminar flame speed in an initially degenerate plasma consisting of \carbon, \oxygen, and \neon[22]. We find that, over a wide range of initial densities and \carbon\ abundances, the flame speed increases roughly linearly with \neon[22] abundance, with the increase being $\approx 30\%$ for $X_{22}= 0.06$, although there are deviations from this rule at lower densities.  These studies are relevant to the initial burning at the near-center of the white dwarf, and at late times where the flame may make a transition to distributed burning. To see how the increase in laminar flame speed changes the density where the burning becomes distributed, we write $\Slam \approx \rho^{\eta}(1 + \xi X_{22})$ and find from our table that at $\rho_{\mathrm{fuel}}
= 7.0\ee{7}\nsp\grampercc$ (the lowest density for which our numerical
scheme converged) and $X_{12}=0.5$, $\eta\approx 1.6$ and $\xi \approx 0.7/0.06$. Recent numerical studies \citep{Zingale2005Three-dimension} find that the Rayleigh-Taylor instability drives turbulence that obeys Kolmogorov statistics, so that $\gibson \propto \Slam^{3}$.  Numerically, we find that the flame width scales roughly as $\Dlam\propto \Slam^{-1.5}$, so solving for $\Dlam/\gibson = 1$ implies that increasing $X_{22}$ from 0 to 0.06 would lower the transition density by $\approx 30\%$. A reduction in the density of this transition will lead to a reduction in the mass of \nickel[56] synthesized \citep{hoflich.khokhlov.ea:delayed}. We
conjecture that if a deflagration-to-detonation occurs, the addition of \neon[22] decreases
the overall mass of Ni-peak elements, in addition to lowering the
isotopic fraction of \nickel[56].

Our results can be improved in several ways. First, the \neon[22] may be partially consumed as $\carbon$ burning gradually heats the core of the white dwarf  \citep{Podsiadlowski2006Cosmological-Im} some $\approx 10^{3}\nsp\yr$ prior to flame ignition. This may further reduce the electron fraction of the white dwarf, but will also change the reaction flows in the flame front.  At low densities the morphology of the flame  becomes more complicated, as the flows responsible for reaching quasi-statistical equilibrium are no longer fast enough to keep up with the carbon burning.  Indeed, at $\rho \le 10^{8}\nsp\grampercc$, the eigenfunction for the flux begins to show two maxima and the flame speed becomes more dependent on the ambient fuel temperature. Further studies with more realistic compositions and at lower ambient densities are ongoing and will be reported in a forthcoming publication.

\acknowledgements 
We thank Diana Hilton, who was supported by the NSF REU program at MSU, for preliminary calculations that motivated this project. We also thank Wolfgang Hillebrandt, Fritz R\"opke, Hendrik Schatz, Michael Wiescher, and Mike Zingale for helpful discussions. This work was supported by the NSF, grant AST-0507456, by the \textbf{J}oint \textbf{I}nstitute for \textbf{N}uclear
\textbf{A}strophysics at MSU under NSF-PFC grant
PHY~02-16783, and by the U.S.\ Dept.\ of Energy via its contract W-7405-ENG-36 to
Los Alamos National Laboratory.

\bibliographystyle{apj}
\bibliography{master}

\end{document}